# Machine learning holography for measuring 3D particle size distribution

Siyao Shao[1, 2], Kevin Mallery[1, 2], Jiarong Hong[1, 2, *]

1.  Saint Anthony Falls Laboratory, 2 3$^{rd}$ AVE SE, University of Minnesota, Minneapolis, MN, USA 55414.
2.  Department of Mechanical Engineering, 111 Church ST SE, University of Minnesota, Minneapolis, MN, USA 55414.

* Email addresses of the corresponding author: jhong@umn.edu

## Abstract

Particle size measurement based on digital holography with conventional algorithms are usually time-consuming and susceptible to noises associated with hologram quality and particle complexity, limiting its usage in a broad range of engineering applications and fundamental research. We propose a learning-based hologram processing method to cope with the aforementioned issues. The proposed approach uses a modified U-net architecture with three input channels and two output channels, and specially-designed loss functions. The proposed method has been assessed using synthetic, manually-labeled experimental, and water tunnel bubbly flow data containing particles of different shapes. The results demonstrate that our approach can achieve better performance in comparison to the state-of-the-art non-machine-learning methods in terms of particle extraction rate and positioning accuracy with significantly improved processing speed. Our learning-based approach can be extended to other types of image-based particle size measurements.

## 1. Introduction

Measurement of particle (e.g. droplets, sediment, bubbles, etc.) size distribution is critical for understanding multiphase flow dynamics in many engineering applications and fundamental research. Some examples are droplet spray generated by pesticide sprayers (Gil and Sinfort 2005), bubbly flows in a chemical reactor (Wang et al. 2014), particle suspensions in fluidized beds (Feng et al. 2017), and pollutant transport in the atmosphere (Stainer et al. 2004). Recently, digital inline holography (DIH) has emerged as a low-cost and compact tool for measuring particle size distributions, particularly in the regime of multiphase flows with low void fraction dispersed phases such as water droplets in cloud (Beals et al. 2015), airborne coal particles (Wu et al. 2014), aerosol generated by breaking wave impingement of oil slicks (Li et al. 2017), drifting droplets from sprays (Kumar et al. 2019), and bubbles in the wake of a ventilated supercavity (Shao et al. 2019a). DIH employs a coherent light source (e.g., laser) and a single camera to record the interference pattern (i.e., hologram) generated from the scattered light of an object and the non-scattered portion of the illumination light source (Katz and Sheng 2010). Traditionally, the hologram is reconstructed through a digital process and the information (i.e. size and location) of the objects within the hologram is extracted through a segmentation process from the reconstructed optical field.

In comparison to other optical-based particle size measurement techniques such as laser diffraction (Eshel et al. 2004), light field imaging (Brücker 2000), and shadowgraphy (Estevadeordal and Goss 2005), DIH can provide high resolution measurements of the 3D distributions of particle size and shape with no assumptions about the shape of the particles like laser diffraction (Eshel et al. 2004). However, the major challenge of DIH lies in the segmentation of the object from the optical fields reconstructed from holograms. Several object segmentation approaches have been proposed in the literature. Tian et al. (2010) used a minimum intensity metric on the object edge in the holograms to determine the object depth and a Gaussian mixture model to segment individual objects from clusters. The method was used for measuring the size distribution of bubbles in a well-mixed water tank with mostly spherically shaped bubbles. A similar approach was adopted by Sentis et al. (2018) for measuring a mixture of diluted bubbles and oil droplets rising in a quiescent water tank. They showed that the bubbles and oil droplets could be discriminated from holograms based on their intensity signatures. However, the performance of the minimum intensity metric is hampered by increasing complexity of the particle field due to increasing particle concentrations and wider ranges of particle sizes. The depth of particles can also be determined through quantification of the level of particle focus using various methods (e.g. Gao et al. 2014, Wu et al. 2014). For example, Gao et al. (2014) used the pixel intensity gradient to quantify sharpness of the particle edge (which determines the particle depth), and measured particle size using the minimum intensity metric. This method has been employed in various applications such as measurements of spherical solid particles in quiescent flow (Gao et al. 2014) and spray measurements in a wind tunnel (Kumar et al. 2019). Furthermore, Wu et al. (2014) applied a wavelet filter to the reconstructed optical field and used the resultant filtered image as the focus metric. They successfully conducted the 3D measurement of irregularly-shaped coal particles in the air. Nevertheless, these types of measurement are only suitable for particles with large sizes relative to the sensor resolution and their performances are susceptible to the noise in holograms such as the virtual image near particle edges (Shao et al. 2019a). Additional particle shape segmentation criterion is often adopted to improve the accuracy of segmentation and localization of the particles from holograms. Particularly, Talapatra et al. (2012) assumed spherical shape in the segmentation and determined the depths of particles based on the pixel intensity gradient calculated from a Sobel filter. Using this approach, Talapatra et al. conducted a measurement of bubble size distribution in a ship wake using holograms captured by a DIH setup towed by the ship and Li et al. (2017) measured the size distribution of droplets generated by a breaking wave impinging on an oil slick. This method must assume that particles have a spherical shape which largely limits its application in measurement tasks, especially for solid particles with irregular shapes (Khanam et al. 2011 and Wu et al. 2014). Recent development by Shao et al. (2019a) has combined a minimum intensity focus metric and wavelet-based focus metric to achieve a 3D distribution of particles with a wide range of sizes. Specifically, the pixels that showed a prominent intensity peak in their longitudinal intensity profile were separated into the small particle group. Other pixels in the holograms were treated as large particles/clusters or background pixels. The large particles were segmented from 2D minimum intensity projection and their depths were determined using a wavelet-based focus metric. This method automatically divides the particles into two groups, largely (but not exclusively) based on their sizes. The Haar wavelet adopted in the wavelet-based focus metric allows accurate focus estimation of single pixels which can be used for estimation of particle 3D orientation. This method has been used for the measurement of bubble size, shape, and orientation in the wake of a ventilated supercavity to estimate instantaneous gas leakage from the cavity. However, this hybrid approach is time-consuming (>5 min per hologram) and requires tuning multiple parameters involved in the segmentation of large particles and the wavelet-based focus metric. In general, previously developed particle segmentation methods in hologram processing are usually time-consuming and sensitive to hologram noise which limits their applications to low concentration particle fields and low background noise. Additionally, these methods usually require fine tuning of parameters to achieve an optimal performance for holograms acquired under different conditions.

Recently, machine learning with deep neural networks (DNNs) has become a powerful tool in object segmentation from noisy images for biomedical and machine vision applications (Barbastathis et al. 2019). For particle analysis, using 2D bright field imaging of bubbly flow, Ilonen et al. (2018) have demonstrated that the adoption of a convolutional neural network (CNN) could yield a higher accuracy in segmenting individual bubbles compared to conventional algorithms like intensity thresholding and watershed segmentation. In hologram processing, the majority of investigations on the application of machine learning have focused on image modality transformations (i.e., transforming hologram reconstruction to commonly used microscopic imaging) (Liu et al. 2019a, and Liu et al. 2019b) and 2D intensity and phase reconstruction of holograms (Rivenson et al. 2018, Wang et al. 2018, and Wang et al. 2019). For single object holograms, the 3D information can be extracted using learning-based regression approaches (Hannel el al. 2019, Ren et al. 2018 and Jaferzadeh et al. 2019). Recent work has employed CNNs in 3D reconstruction of tracer fields (Shimobaba et al. 2019, Shao et al. 2019b). Particularly, Shao et al. 2019(b) used a modified U-net architecture with added residual connections and Swish activation function, and a Huber loss function and total variation (TV) regularized mean square error (MSE) loss function for particle field reconstruction. Hologram preprocessing and transfer learning were also employed to minimize the training cost. Compared to a prior machine learning approach (Shimobaba et al. 2019) and state-of-art inverse reconstruction (Mallery and Hong 2019), the newly proposed method has demonstrated higher particle extraction rate and positioning accuracy for both synthetic and experimental holograms, especially with high particle concentrations up to 0.06 particle per pixel. However, to our best knowledge, machine learning has only been applied for characterization of size and shape of particles for simple cases such as a very dilute sample of monodisperse colloidal particles (Yevick et al. 2014) or on synthetic holograms of single white blood cells (Chneider et al. 2016).

In the present paper, we propose a learning-based 3D particle measurement method using holograms. This approach aims to address the issues related to particle segmentation from reconstructed holograms like complex segmentation criteria and tuning of parameters. Section 2 of this paper provides the general approach including the machine learning architecture, hologram preprocessing and labeling, and loss functions. In Section 3, we present the assessment of this approach using synthetic holograms following by Sections 4 and 5 using real holograms. Lastly, Section 6 is a summary and discussion of the presented results.

## 2. Methodology

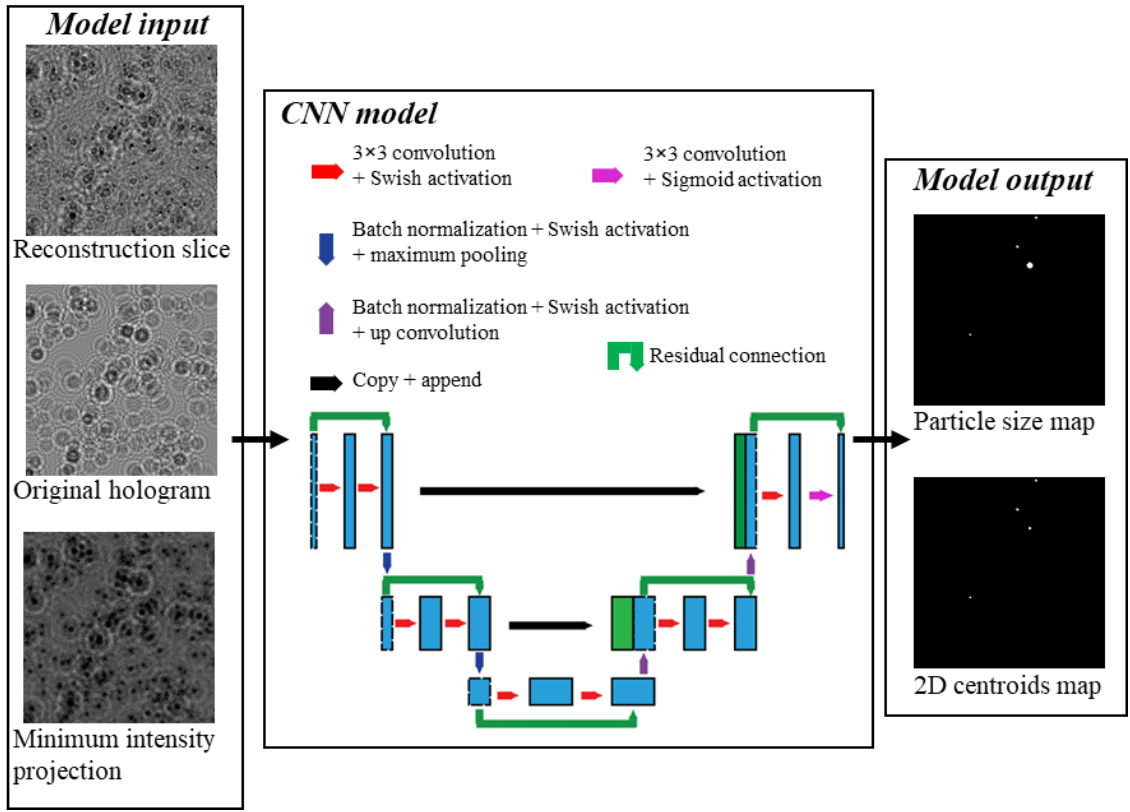

Fig.1. Schematic illustrating the proposed machine learning method for particle segmentation from holograms.

The machine learning model used in the current study is adapted from the modified U-net architecture of Shao et al. (2019b) (Fig.1). As discussed in the introduction, our architecture adopts residual connections to improve training speed and accuracy of the model prediction. The Swish (Sigmoid-weighed Linear Unit) activation function replaces the commonly-used ReLU (rectified linear unit), improving the performance when training with sparse targets like particle fields. The training input has three channels: the image reconstructed from hologram at a particular longitudinal location, the original hologram, and a minimum intensity projection from the reconstructed optical field of the hologram (Fig. 1). To produce these channels, the original hologram is first convolved with a Rayleigh-Sommerfeld kernel to get a 3D complex optical field, $u_p(x, y, z)$ (Eqn. 1).

$$u_p(x, y, z) = \mathcal{F}^{-1}\left\{\mathcal{F}[I_k(x, y) \times \mathcal{F}[\frac{1}{j\lambda\sqrt{x^2+y^2+z^2}}\exp[jk(\sqrt{x^2+y^2+z^2})]]]\right\} \quad (1)$$

The $I_k$ refers to the enhanced hologram captured in $(x, y)$ planar space and $z$ corresponding to each longitudinal location. The $\lambda$ in the above equation is the wavelength of the illumination beam, $k$ is the wave number and $\mathcal{F}$ is the fast Fourier transform operator. The pixel intensity at each location can be calculated as the magnitude of complex intensity values from $u_p(x, y, z)$ and the minimum intensity projection image is generated by projecting the longitudinal minimum intensity of each pixel onto a plane. The hologram preprocessing used in the present approach reduces the need of the machine learning model to fully learn the required physics of hologram formation during training (Barbastathis et al. 2019). As for the training target, a binary image consisting of only in-focus (at the reconstruction depth) particles with their true shapes is employed as the particle size map (Fig. 1). Additionally, a 2D particle centroids map comprised of particle centroid regions (each 2×2 pixels) is used for determination of particle location and helps remove false detections (i.e., ghost particles) from the prediction.

The training loss for the particle size map channel is a modified generalized dice loss (GDL, Eqn. 2) which is capable of multiclass classification. Here, $N = 1$ since only in-focus particle pixels are classified using the model. As shown in Eqn. (2), this loss function first calculates the ratio of overlapped areas of class $n$ pixels in the ground truth ($X_n$) and model prediction ($Y_n$) and a modified area of union of $X_n$ and $Y_n$. The training loss is this ratio subtracted from 1 and $\delta$ is a relaxation factor to prevent division by zero during training (set as 0.02 in our case). As noted by Sudre et al. (2017), GDL has very good performance on rare species in classification problems (e.g., in-focus particles on each reconstruction plane). We further modify the measure of the area of union in the dominator as the sum of *L2* norms of $Y_n$ and $X_n$ instead of their *L1* norms in Sudre et al. (2017). This modification improves the training stability as suggested in Ding et al. (2006).

$$L = 1 - 2 \left(\sum_{n=1}^{N} Y_n X_n + \delta\right) / \left(\sum_{n=1}^{N} Y_n^2 + X_n^2 + \delta\right) \quad (2)$$

For the particle 2D centroids channel, we adopt a total variation (TV) regularized mean square error (MSE) loss (Eq. 3). As shown in Eqn. 4, TV is the sum of first-order gradients over the image (size $Nx \times Ny$). This loss function, as suggested by Shao et al. (2019b), could force the model to generate a prediction with clean background and remove the false detections (i.e., ghost particles). Note that in our case, the parameter $\alpha$ that controls the smoothness of model outputs is set as 0.0001. Based on our tests, increasing $\alpha$ will degrade the accuracy of the results since TV acts as a filter and may blur the model output with high $\alpha$.

$$L = (1 - \alpha)(\|Y - X\|_2^2) + \alpha \times [TV(Y)]^2 \quad (3)$$

$$TV(Y) = \sum_{i=1}^{Nx} \sum_{j=1}^{Ny} \sqrt{(Y_{i,j} - Y_{i-1,j})^2 + (Y_{i,j} - Y_{i,j-1})^2} \quad (4)$$

The post-processing to obtain 3D particle shape and size distribution requires first finding the candidate particles from the particle size channel using an averaged intensity of 0.3 of each connected object. For each candidate particle, within consecutive ±5 reconstruction planes, the particle 3D centroids are determined by using the location of highest intensity object from 2D centroids outputs from the same region determined by the bounding boxes of the candidate particles. On the reconstruction plane found from the previous step, a binarization using threshold of 0.3 is conducted in the particle bounding boxes to find particle regions. For the ROI without any particle centroids from particle location channel outputs, we treat the candidate particles as false detections. For the ROI with multiple centroids on the same reconstruction plane, a marker-controlled watershed segmentation is conducted on the binary image from the size channel using the particle centroids from particle location channel as markers (Gonzalez et al. 2009). The calculation of particle size (i.e., area equivalent diameter $d = \sqrt{4A\pi}$, where $A$ is the area occupied by each segmented particles), and particle shape (i.e., eccentricity) follows the approach provided in Karn et al. (2015). Compared to the previous particle segmentation methods, our post-processing does not require any preset parameters other than an intensity threshold unlike previous methods (e.g., Talapatra et al. 2012, Shao et al. 2019a). In addition, our approach can segment particles that are occluded or overlapped by applying the segmentation on each reconstruction plane rather than on a 2D projection of the volume. This allows for higher concentrations compared to prior methods (e.g., Wu et al. 2014 and Sentis et al. 2018).

## 3. Assessment of proposed approach using synthetic holograms

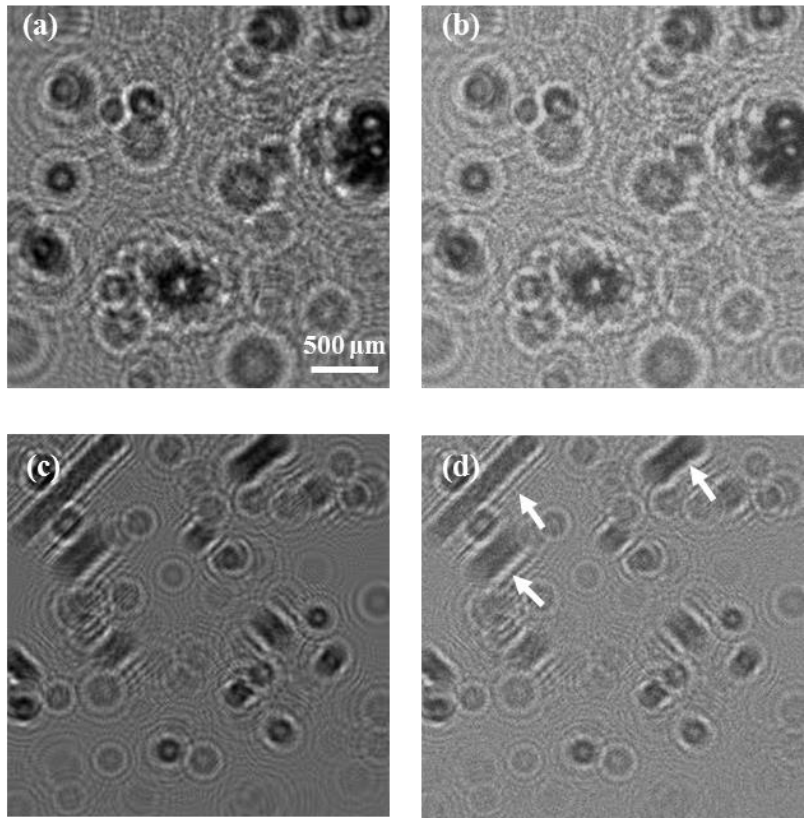

Fig.2. A sample of spherical particle holograms synthesized for assessment before (a) and after (b) data augmentation and poly-shape (spherical and elliptical) particle hologram before (c) and after (d) data augmentation. Examples of elliptical particles are labeled by the arrows in (d).

The machine learning approach is firstly assessed through synthetic holograms. The hologram synthesis follows the approach described in the literature (e.g. Zhang et al. 2006, Gao 2014) Holograms consisting of pure spherical particles and holograms consisting of both spherical and elliptical particles are used in this assessment (Fig. 2). Each hologram has 200 particles following a log-normal distribution of equivalent diameter, with particle size ranging from 0.02 to 0.83 mm

(the average particle size is 0.04 mm with 0.03 mm standard deviation). The particles are uniformly distributed in depth with a distance to the image plane between 20 mm and 40 mm. In the holograms consisting of elliptically shaped particles, all the synthesized particles with equivalent diameters smaller than 0.2 mm are set as spherical and all other particles have a random distributed eccentricity from 0 (spherical) to 0.85 (elongated elliptical shapes). The synthetic holograms are generated using 11.1 μm/pixel resolution under 532 nm laser illumination and each hologram has a size of 256×256 pixels. All the holograms are augmented by applying a Gaussian filter and adding Gaussian noise (Fig. 2). As suggested by Frid-Adar et al. (2018), data augmentation significantly increases the diversity of data available for training without actually collecting or synthesizing new samples. As a result, the model is less likely to overfit the data when it is trained with an augmented dataset. The model is trained on a dataset of 100 holograms (20000 particles)) with only spherical particles for 40 epochs. The preprocessing uses a reconstruction depth step size of 40 μm. The training uses TensorFlow 2.0 on an Nvidia RTX 2080Ti GPU. The Adam optimizer is employed in the training with default settings (Kingma and Ba 2014). The total training time is about 15 hrs. After training, the trained model is tested on both particle holograms with pure spherical particles and holograms containing elliptical particles. Each test dataset consists of 100 holograms (20000 particles total). It is worth noting that although the range of particle size is similar for training and testing datasets, we use a different particle mean diameter (0.03 mm) and standard deviation (0.02 mm) to conduct the tests.

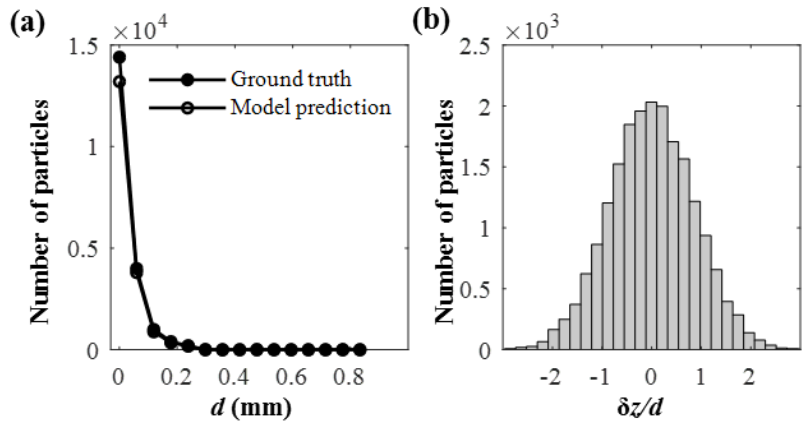

Fig.3. (a) A comparison of the ground truth and measured distribution from the proposed machine learning method for mono-shape particle dataset and (b) the histogram of longitudinal positioning error.

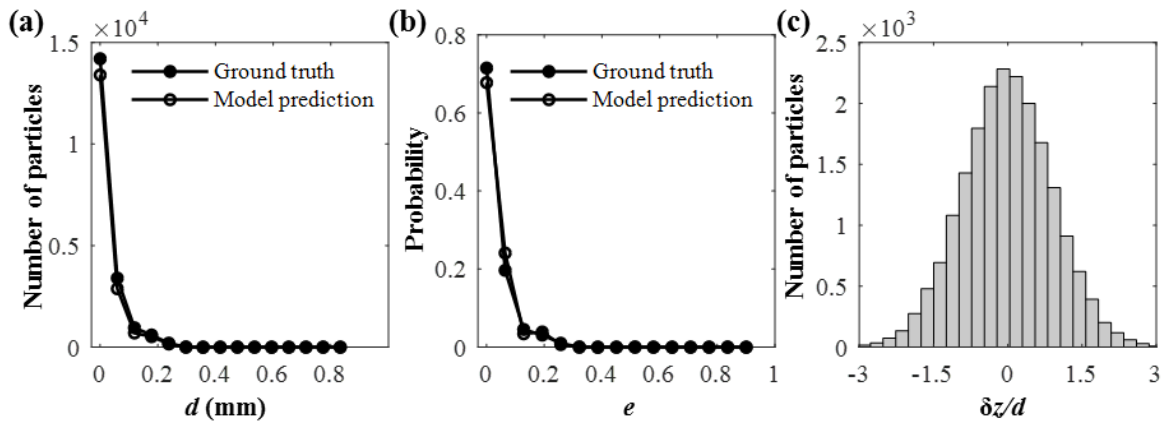

Fig.4. (a) A comparison of the ground truth and measured distribution from the proposed machine learning method for poly-shape particle dataset and (b) eccentricity of all the detected particles compared to ground truth and (c) the histogram of longitudinal positioning error.

As shown in Fig. 3 (a), the resultant particle size distribution agrees well with the ground truth over the entire particle size range for the dataset consisting of only spherical particle shapes. The machine learning approach has a detection rate of 92% with over 90% detected particles having a positioning error in longitudinal direction ($\delta z$) less than a particle diameter (Fig. 3b). This detection rate from the present method is slightly better than the result from same dataset using conventional method proposed by Shao et al. (2019a) which has a detection rate of 91.2 %. The test results from holograms containing of elliptical particles also exhibit good agreement of particle size distribution (Fig. 4a). Additionally, the eccentricity calculated from the detected particles agrees well with the input values despite a slightly lower percentage of spherical particles from the model prediction (Fig 4b). We suggest such mismatching is due to the undetected small particles since small particles (<0.2 mm) are set as spherical in the synthetic dataset. The comparison demonstrates that our machine learning approach is less dependent of the particle shape variation compared to the methods using shape as segmentation criterion such as Hough transform (Talapatra et al. 2012). This characteristic substantially enhances the flexibility of a model trained on one dataset when applied to new holograms. Lastly, the overall detection rate for the dataset with elliptical particles is about 91% with over 90% detected particles having longitudinal errors less than 1.5 of the particle diameter. This degraded particle positioning accuracy compared to the previous case is primarily due to the increased noise level in the particle position channel output with elongated particles.

## 4. Assessment of proposed approach with labelled experimental data

The proposed method is assessed using experimental holograms particles suspended in water. We use 20 µm, 100 µm and 1 mm particles dispersed in water in a cuvette (10 mm in depth). The experimental holograms (Fig. 5a) are recorded on a NAC HX-5 camera using a 105 mm focal length Nikon lens (magnification at 1:1) with a collimated 532 nm diode laser. The experimental data is manually labeled by selecting the in-focus particles on each reconstruction plane. The centroids and size of each particle are measured using the Hough transform. The training dataset consists of 40 256×256-pixel holograms with each hologram reconstructed to 500 slices. As shown in Fig. 5(c), the measured particle size distribution from a 100-hologram test dataset using the model trained by experimental holograms match well to the ground truth. The machine learning approach has a detection rate of 94.6% with over 90% of detected particles having a positioning error in longitudinal direction ($\delta z$) less than a particle diameter (Fig. 5d). The testing results of the same test set using the non-machine-learning method (Shao et al. 2019a) has shown a detection rate of 88.7% with over 60% of detected particles having a positioning error $z$ direction less than a particle diameter. In general, our proposed learning-based approach has demonstrated noticeable improvement of detection rate and positioning accuracy compared to non-machine-learning method, though the cost of manually labeling ground truth is still high (>30 min per hologram).

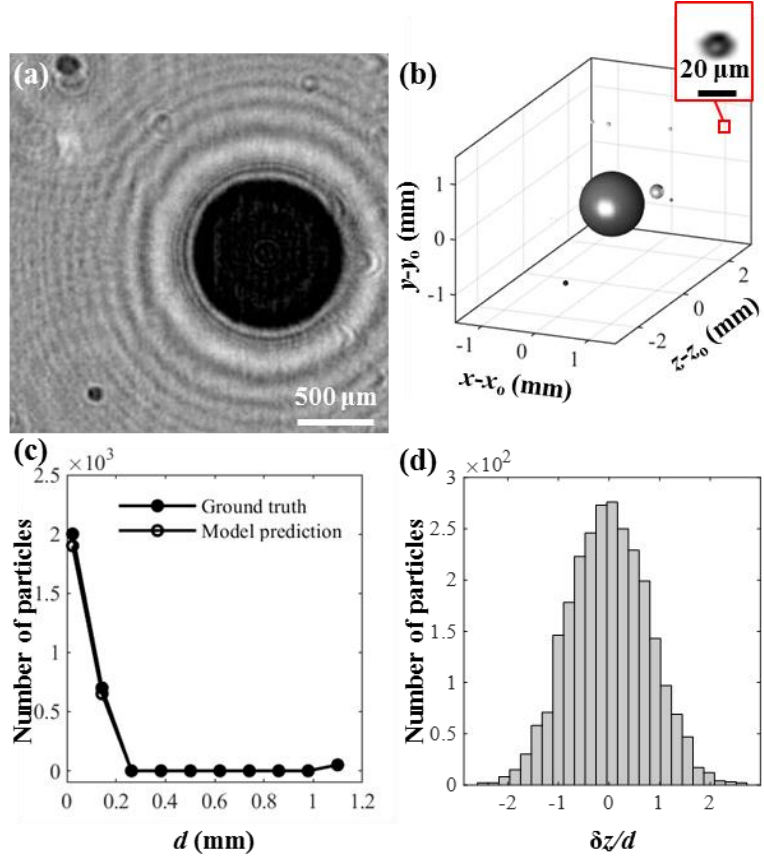

Fig. 5. (a) A sample particle hologram from the experimental data and (b) corresponding rendering of 3D particle distribution from manual labeling. Inset is a magnified view of the red rectangle shown on the plot to depict the small particles captured in the holograms. (c) A comparison of the ground truth and measured distribution from proposed machine learning method and (d) the histogram of the longitudinal positioning error.

## 5. Assessment of proposed approach with bubble holograms

Lastly, the proposed method is assessed through measurement of bubble size distribution in the wake of a ventilated supercavity. The holograms are recorded with a high speed DIH setup comprised of a 532 nm continuous diode laser, beam expansion optics, and a Photron APS-RX high speed camera with a Nikon 105 mm imaging lens. The pixel resolution of the hologram sequence is 51 μm/pixel and the bubbles are dispersed in water. The details of the experiments can be found in Shao et al. (2019a). The reconstruction volume is from 150 mm to the image plane to 250 mm to the image plane with spacing at 0.25 mm. The ground truth on the hologram sequence is established using the method provided in Shao et al. (2019a). A sample hologram of this dataset is shown as Fig. 6(a) and its 3D reconstructed bubble volume is shown as Fig. 6(b). Note that the training is on 80 cropped regions of 256×256 pixel from each hologram random selected from the dataset, but the testing is conducted on a separate set of 50 randomly-selected full-size holograms (i.e., 512×372 pixels) using the model trained from cropped holograms. Such capability is associated with the U-net architecture employed in our machine learning model, which can be directly applied to images of arbitrary size (regardless of the training set image size) owing to the absence of densely connected layers (Ronneberger et al. 2015). The training lasted for 14 hours

(i.e., 20 epochs) using the same hardware and optimizer as the cases presented in Section 3 and 4. As depicted in Fig. 6(c) and Fig. 6(d), the measured bubble size distribution has a good agreement with the ground truth with a slightly higher longitudinal positioning error compared to the synthetic cases in Section 3. We suggest such reduced positioning error is primarily due to lower resolution and inclusion of irregular shaped bubbles in the holograms. It is worth noting the performance of the machine learning model in this assessment is limited by the accuracy of the conventional hologram reconstruction and particle segmentation approach used for building ground truth for the training set. Nevertheless, the machine learning method is 10X faster than non-machine-learning method through traditional reconstruction and segmentation of the holograms. This method is suitable for applications like batch processing a large number of holograms where it is difficult to acquire training ground truth through experiments or hologram synthesis. It is worth noting that the processing speed of our approach can be further improved through additional optimization of the pre- and post-processing steps using GPU, allowing potentially a real-time characterization of particle distribution.

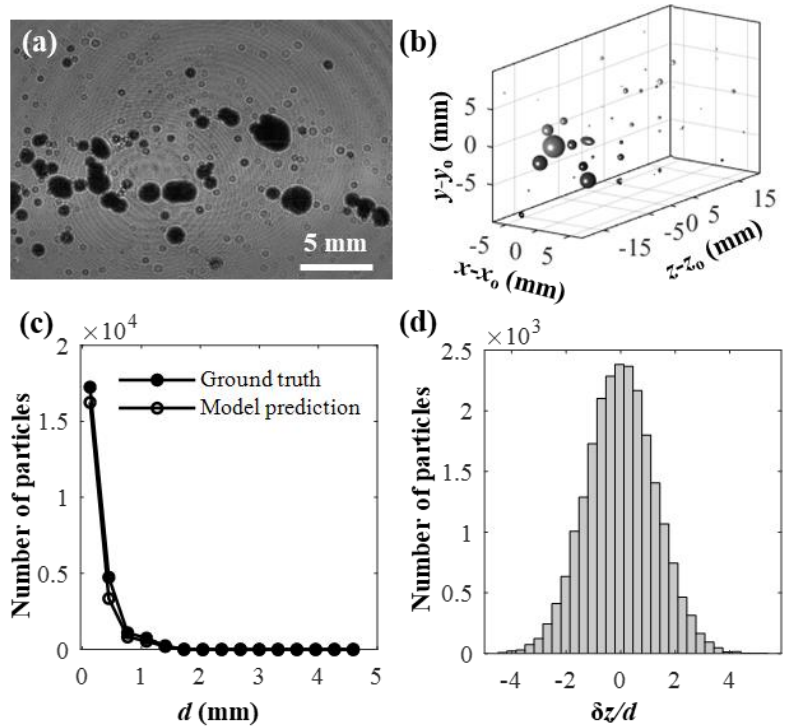

Fig.6. (a) A sample of enhanced bubble hologram used for testing and its 3D bubble distribution reconstructed using the method proposed in Shao et al. (2019a) (b). (c) A comparison of the ground truth and measured distribution from proposed machine learning method using the model trained with the ground truth obtained using the method from Shao et al. and the longitudinal positioning error normalized by particle equivalent diameter (d).

## 6. Summary and discussion

In this paper, we propose a machine learning approach for 3D measurement of particle size distribution using digital holography. A modified U-net architecture is employed as the machine learning model to directly extract particle size from holograms. Three channels, original hologram, hologram reconstructed to each longitudinal location, and minimum intensity projection in longitudinal direction of the reconstructed optical field, are used as the model input. The preprocessing of the holograms reduces the need of the model to fully learn the physics required for the hologram formation. The model outputs are one channel encoded only in focus particle shapes corresponding to each reconstruction depth and one channel encoded in focus particle 2D centroid regions. The output channel with particle centroid regions is used for determining particle locations and helps remove false detections from the particle size channel. The training of the machine learning model uses a modified generalized dice loss for particle size channel and a total variation regularized mean squared error loss for 2D centroids channel. Our approach has been assessed through synthetic holograms of spherical and elliptical particles, manually-labeled experimental holograms, and water tunnel bubbly flowdata containing particles of different shapes. The assessment has demonstrated that the machine learning approach has even better performance compared to the state-of-the-art non-machine-learning methods (e.g., Shao et al. 2019a), in terms of extraction rates and positioning accuracies for both synthetic and experimental data with significant improvement in processing speed.

Overall, we have developed a novel learning-based approach for measurement of 3D particle distribution based on holographic imaging with improved particle extraction rate, positioning accuracy and processing speed. This method can be used for holographic-based 3D imaging and size analysis of various particles including droplets, sprays, bubbles, sediments and other complex-shaped objects such as marine microorganisms. Additionally, the proposed machine learning framework can be readily extended to other types of image-based size measurement tasks such as brightfield imaging (Brücker 2000, Ilonen et al. 2018), shadowgraph imaging (Karn et al. 2015), defocusing imaging (Roma et al. 2014), etc.

Although the proposed approach has shown substantial speed improvement, we suggest that further optimization on pre- and post-processing steps of our approach will likely to result in significant boost of processing speed, potentially enabling *in situ* real-time analysis of particle size with holography. Nevertheless, it is worth noting that our current machine learning approach still relies on labeled experimental data for training to achieve high accuracy, which is usually time consuming and labor intensive. Future work will focus on improving our learning model with training based on synthetic holograms, which eliminates the need of collecting training data through manually-labeling or non-machine-learning methods.

## Acknowledgements

This work is supported by the Office of Naval Research (Program Manager, Dr. Thomas Fu) under grant No. N000141612755 and the start-up funding received by Prof. Jiarong Hong from University of Minnesota.

## References

Barbastathis, G., Ozcan, A., Situ, G., 2019. On the use of deep learning for computational imaging. Optica 6, 921-943.

Beals, M. J., Fugal, J. P., Shaw, R. A., Lu, J., Spuler, S. M., Stith, J. L., 2015. Holographic measurements of inhomogeneous cloud mixing at the centimeter scale. Sci. 350(6256), 87-90.

Brücker, C., 2000. PIV in two-phase flows. von Karman Institute for fluid dynamics, Lecture Series, 1.

Chneider, B. S., Ambre, J. D., 2016. Fast particle characterization using digital holography and neural networks. Appl. Opt. 55(1): 133-139.

Ding. C., Ding. Z., He. X., Zha.H., R1-PCA: rotational invariant L1-norm principal component analysis for robust subspace factorization. in Proc. of 23rd ICML 2006-Pittsburg, (ACM, 2006), pp. 281-288.

Eshel, G., Levy, G. J., Mingelgrin, U., Singer, M. J., 2004. Critical evaluation of the use of laser diffraction for particle-size distribution analysis. Soil Sci. Soc. Am. J. 68(3), 736-743.

Estevadeordal, J., Goss, L., PIV with LED: particle shadow velocimetry (PSV) technique, in: 43rd AIAA aerospace sciences meeting and exhibit, 2005.

Feng, R., Li, J., Cheng, Z., Yang, X., Fang, Y., 2017. Influence of particle size distribution on minimum fluidization velocity and bed expansion at elevated pressure. Powder Technol. 320, 27-36.

Frid-Adar, M., Klang, E., Amitai, M., Goldberger, J., Greenspan, H., Synthetic data augmentation using GAN for improved liver lesion classification. In 2018 IEEE 15th international symposium on biomedical imaging (ISBI 2018), pp. 289-293.

Gao, J., 2014. Development and applications of digital holography to particle field measurement and in vivo biological imaging. PhD diss., Purdue University.

Gao, J., Guildenbecher, D.R., Engvall, L., Reu, P.L., Chen, J., 2014. Refinement of particle detection by the hybrid method in digital in-line holography. Appl. Opt. 53(27), 130-138.

Gil, Y., Sinfort, C., 2005. Emission of pesticides to the air during sprayer application: A bibliographic review. Atmos. Environ. 39(28), 5183-5193.

Gonzalez, R.C., Woods, R.E., Eddins, S.L., 2009. Digital Image Processing Using MATLAB®. Gatesmark Publishing.

Hannel, M. D., Abdulali, A., O'Brien, M., Grier, D. G., 2018. Machine-learning techniques for fast and accurate feature localization in holograms of colloidal particles. Opt. Express 26, 15221-15231.

Ilonen, J., Juránek, R., Eerola, T., Lensu, L., Dubská, M., Zemčík, P., Kälviäinen, H., 2018. Comparison of bubble detectors and size distribution estimators. Pattern Recognit. Lett.101, 60-66.

Jaferzadeh, K., Hwang, S. H., Moon, I., & Javidi, B. (2019). No-search focus prediction at the single cell level in digital holographic imaging with deep convolutional neural network. Biomed. Opt. Express 8, 4276-4289.

Karn, A., Ellis, C., Arndt, R. E., Hong, J., 2015. An integrative image measurement technique for dense bubbly flows with a wide size distribution. Chem. Eng. Sci. 122, 240-249.

Katz, J., Sheng, J., 2010. Applications of holography in fluid mechanics and particle dynamics. Annu. Rev. Fluid Mech. 42, 531-555.

Khanam, T., Rahman, M.N., Rajendran, A., Kariwala, V., Asundi, A.K., 2011. Accurate size measurement of needle-shaped particles using digital holography. Chem. Eng. Sci. 66(12), 2699-2706.

Kingma. D. P., Ba. J., 2014 Adam: a method for stochastic optimization axXiv: 1412.6980.

Kumar, S. S., Li, C., Christen, C. E., Hogan Jr, C. J., Fredericks, S. A., Hong, J., 2019. Automated droplet size distribution measurements using digital inline holography. J. Aerosol Sci. 137, 105442.

Li, C., Miller, J., Wang, J., Koley, S. S., Katz, J., 2017. Size distribution and dispersion of droplets generated by impingement of breaking waves on oil slicks. J. Geophys. Res. Oceans 122(10), 7938-7957.

Liu, T., de Haan, K., Rivenson, Y., Wei, Z., Zeng, X., Zhang, Y., Ozcan, A., 2019. Deep learning-based super-resolution in coherent imaging systems. Sci. Rep. 9(1), 3926.

Liu, T., Wei, Z., Rivenson, Y., de Haan, K., Zhang, Y., Wu, Y., Ozcan, A., 2019. Deep learning-based color holographic microscopy. J. Biophotonics, e201900107.

Mallery, K., Hong, J., 2019. Regularized inverse holographic volume reconstruction for 3D particle tracking. Opt. Express 27, 18069-18084.

Ren, Z., Xu, Z., Lam, E. Y., 2018. Learning-based nonparametric autofocusing for digital holography. Optica, 5(4), 337-344.

Rivenson, Y., Zhang, Y., Günaydin, H., Teng, D., Ozcan, A., 2018. Phase recovery and holographic image reconstruction using deep learning in neural networks. Light Sci. Appl. 7, 17141.

Roma, P.M.S., Siman, L., Amaral, F.T., Agero, U., Mesquita, O.N., 2014. Total three-dimensional imaging of phase objects using defocusing microscopy: application to red blood cells. Appl. Phys. Lett. 104, 251107-1-251107-5.

Ronneberger, O., Fischer, P., & Brox, T., 2005 U-net: Convolutional networks for biomedical image segmentation. In International Conference on Medical image computing and computer-assisted intervention (pp. 234-241). Springer, Cham.

Sentis, M.P., Onofri, F.R., Lamadie, F., 2018. Bubbles, drops, and solid particles recognition from real or virtual photonic jets reconstructed by digital in-line holography. Opt. Lett. 43(12), 2945-2948.

Shao, S., Li, C., Hong, J. 2019a. A hybrid image processing method for measuring 3D bubble distribution using digital inline holography. Chem. Eng. Sci. 207, 929-941.

Shao, S., Mallery, K., Kumar, S., Hong, J., 2019b. Machine Learning Holography for 3D Particle Field Imaging. arXiv preprint arXiv:1911.00805.

Shimobaba, T., Takahashi, T., Yamamoto, Y., Endo, Y., Shiraki, A., Nishitsuji, T., Hoshikawa, N., Kakue, T., Ito, T., 2019. Digital holographic particle volume reconstruction using a deep neural network. Appl. Opt. 58, 1900-1906.

Stanier, C. O., Khlystov, A. Y., Pandis, S. N., 2004. Ambient aerosol size distributions and number concentrations measured during the Pittsburgh Air Quality Study (PAQS). Atmos. Environ. 38(20), 3275-3284.

Sudre, C.H., Li, W., Vercauteren, T., Ourselin, S., Cardoso, M.J., 2017. Generalised dice overlap as a deep learning loss function for highly unbalanced segmentations. In Deep learning in medical image analysis and multimodal learning for clinical decision support (pp. 240-248). Springer, Cham.

Talapatra, S., Sullivan, J., Katz, J., Twardowski, M., Czerski, H., Donaghay, P., Hong, J., Rines, J., McFarland, M., Nayak, A.R., Zhang, C., Application of in-situ digital holography in the study of particles, organisms and bubbles within their natural environment, in: Ocean Sensing and Monitoring IV, 2012.

Tian, L., Loomis, N., Domínguez-Caballero, J. A., Barbastathis, G., 2010. Quantitative measurement of size and three-dimensional position of fast-moving bubbles in air-water mixture flows using digital holography. Appl. Opt. 49(9), 1549-1554.

Wang, A., Marashdeh, Q., Fan, S., 2014. ECVT imaging of 3D spiral bubble plume structures in gas-liquid bubble columns. Can. J. Chem. Eng. 92, 2078-2087.

Wang, H., Lyu, M., Situ, G., 2018. eHoloNet: a learning-based end-to-end approach for in-line digital holographic reconstruction. Opt. Express 26(18), 22603-22614.

Wang, K., Dou, J., Kemao, Q., Di, J., Zhao, J., 2019. Y-Net: a one-to-two deep learning framework for digital holographic reconstruction. Opt. Lett. 44(19), 4765-4768.

Wu, Y., Wu, X., Yang, J., Wang, Z., Gao, X., Zhou, B., Chen, L., Qiu, K., Gerard, G., Chen, K., 2014. Wavelet-based depth-of-field extension, accurate autofocusing, and particle pairing for digital inline particle holography. Appl. Opt. 53(4), 556-564.

Yevick, A., Hannel, M., Grier, D. G., 2014. Machine-learning approach to holographic particle characterization. Opt. Express 22(22), 26884–26890.

Zhang, Y., Shen, G., Schröder, A., Kompenhans, J., 2006. Influence of some recording parameters on digital holographic particle image velocimetry. Opt. Eng. 45(7), 075801.